\documentstyle[preprint,aps]{revtex}
\begin{document}
\draft

\preprint{\begin{minipage}[b]{1.14in}
          CU-TP-885\\
          hep-ph/9804234
          \end{minipage}}
\vspace{0.2in}

\title{Jet broadening in deeply inelastic scattering
\thanks{This work was partially supported by the U.S. Department 
of Energy under contract No. DE-FG02-93ER40764.}}
\author{Xiaofeng Guo}
\address{Department of Physics, Columbia University \\
         New York, NY 10027, USA}
\date{April 6, 1998}
\maketitle

\begin{abstract}
In deeply inelastic lepton-nucleus scattering (DIS), 
the average jet transverse momentum 
is broadened because of multiple scattering in the nuclear medium. 
The size of jet broadening is proportional to the 
multi-parton correlation functions inside nuclei. We show that 
at the leading order, 
jet broadening in DIS and nulcear enhancement in di-jet momentum imbalance  
and Drell-Yan $\langle q_T^2 \rangle $ share the same four-parton 
correlation functions. We argue that jet broadening in DIS provide  
an indenpendent measurement of the four-parton correlation functions and  
a test of QCD treatment of multiple scattering.
\end{abstract} 
\vspace{0.2in}
\section{Introduction}

When a parton propagates through a nuclear matter, the average 
transverse momentum may be broadened because of multiple scattering 
of the parton. Nuclear dependence of such broadening provides an 
excellent probe to study QCD dynamics beyond the single 
hard-scattering picture.
Reliable calculation of multiple scattering in QCD perturbation theory 
requires to extend the factorization theorem \cite{Factorization} 
beyond the leading power.  Qiu and Sterman showed that the factorization 
theorem for hadron-hadron scattering holds at 
the first-nonleading power in momentum transfer \cite{QS}, 
which is enough to study double scattering processes in hadronic 
collisions.
According to this generalized factorization theorem, the double 
scattering contribution 
can be expressed in terms of universal four-parton (twist-four)
correlation functions. The  
predictions of the multiple scattering effects rely on the accurate 
information of the four-parton correlation functions. Various 
estimate of the size of these four-parton correlation functions were 
derived recently \cite{LQS2,YuriD,Baieretal}. 
It is the purpose of this paper to show that  
jet broadening in deeply inelastic lepton-nucleus scattering (DIS)  
can be used as an independent process to test the previous 
estimate of the four-parton correlation functions.

The four-parton (twist-4) correlation functions are as fundamental as 
the normal twist-2 parton distributions. While twist-2 parton 
distributions have 
the interpretation of the probability distributions to find a parton
within a hadron, four-parton correlation functions provide information 
on quantum correlations of multi-partons inside a hadron.  
Just like normal twist-2 parton distributions,
multi-parton correlation functions are non-perturbative.  QCD
perturbation theory cannot provide the absolute prediction of these
correlation functions.  But, due to factorization theorems,
these correlation functions are universal; and can be measured in some
processes and be tested in other processes.  

Luo, Qiu and Sterman (LQS) calculated the nuclear dependence of di-jet 
momentum imbalance in photon-nucleus collision in terms of 
nuclear four-parton
correlation functions \cite{LQS2},
\begin{mathletters}
\begin{eqnarray}
T_{q/A}(x) &=&
\int \frac{dy^-}{2\pi}\, e^{ixp^+y^-}\
\frac{dy_1^-dy_2^-}{2\pi}  
\theta (y_1^--y^-) \,\theta (y_2^-)\, 
\nonumber \\
&\ & \times \, \frac{1}{2}\,
\langle p_A |
 \bar{\psi}_q(0)\, \gamma^+ \,F_{\sigma}^{\ +}(y_{2}^{-})\,
          F^{+\sigma}(y_{1}^{-})\, \psi_q(y^{-}) | p_{A}\rangle \ ;
\label{Ta}
\end{eqnarray}
and
\begin{eqnarray}
T_{g/A}(x) &=&
\int \frac{dy^-}{2\pi} \, e^{ixp^+y^-}\
\frac{dy_1^-dy_2^-}{2\pi}  
\theta (y_1^--y^-) \,\theta (y_2^-)\, 
\nonumber \\
&\ & \times \, \frac{1}{xp^+}\,
\langle p_A |
 F_{\alpha}^{\ +}(0) \,F_{\sigma}^{\ +}(y_{2}^{-})\,
          F^{+\sigma}(y_{1}^{-})\, F^{+\alpha}(y^{-}) | p_{A}\rangle \ .
\label{Tb}
\end{eqnarray}
\end{mathletters}
By comparing the operator 
definitions of these four-parton correlation functions and
 the definitions of the normal twist-2 parton distributions,  
LQS proposed the following model  \cite{LQS2,LQS1}:
\begin{equation}
T_{f/A}(x)=\lambda^2 A^{4/3} \phi_{f/N}(x) \ ,
\label{TiM}
\end{equation}
where $\phi_{f/N}(x)$ with $f=q,\bar{q},g$ are the normal twist-2 parton 
distribution of a nucleon, and $\lambda$ is a free parameter 
to be fixed by experimental data.
Using the Fermilab E683 data on di-jet momentum
imbalance, LQS estimated the size of the relevant four-parton
correlation functions to be of the order  $\lambda^2 \approx 
0.05\sim 0.1$ GeV$^2$ \cite{LQS2}.  

The nuclear enhancement of the average Drell-Yan transverse momentum, 
$\Delta \langle q_T^2 \rangle $,  
also depends on the similar four-parton correlation functions  
\cite{YuriD}. 
However, Fermilab and CERN data on nuclear dependence of the Drell-Yan 
$\Delta \langle q_T^2 \rangle $ prefer a much smaller size of the 
four-parton correlation functions \cite{YuriD,E772,NA10}. 
Actually, as we will show below, the Drell-Yan data favor the four-parton 
correlation functions about five times smaller than what was 
extracted from the di-jet data.  
This discrepancy may result from different higher order contribution to 
the Drell-Yan $\Delta \langle q_T^2 \rangle $ and the di-jet momentum
imbalance. We have pure initial-state multiple scattering for 
the Drell-Yan process, while pure final state multiple scattering 
for a di-jet system. It is necessary to study the high order
corrections to these two observables in order to test QCD treatment of 
multiple scattering. It was argued in Ref.~\cite{Guo,GQS}  that the high 
order contribution to the Drell-Yan nuclear enhancement is important. 
Meanwhile, it is also 
important and necessary to find different observables that depend on 
the same four-parton correlation functions.

The jet broadening in deeply inelastic lepton-nucleus scattering
provides an independent test of the size of the four-parton correlation 
functions. At the leading order in deeply inelastic scattering, 
a quark scattered of the virtual 
photon forms a jet (known as the current jet) in the final state. 
Because of the multiple interaction when the scattered 
parton propagates through the nuclear matter,  
the transverse momentum of the final state jet is  
broadened in deeply inelastic lepton-nucleus scattering \cite{strikman}. 
We show below that the size of 
the jet broadening is directly proportional to the four-parton 
correlation functions. These four-parton correlation functions 
are the same as those appeared in the nuclear enhancement of 
the di-jet momentum imbalance. 
Current data for the di-jet momentum imbalance
and the Drell-Yan $\Delta \langle q_T^2 \rangle$  
provide two different sizes of 
the four-parton correlation functions. Using these data, 
without additional parameters, we predict 
a range of the jet broadening in DIS. Existing data from Fermilab E665 
and future HERA nuclear beam data 
can provide a direct test of QCD treatment of multiple scattering. 

The rest of this paper is organized as follows.  In Sec.~II, we derive
the nuclear enhancement of the Drell-Yan $\langle q_T^2 \rangle$ 
by using the same
method used by LQS to derive the di-jet momentum imbalance,
and show that our result is consistent with that presented 
in Ref.~\cite{YuriD}.  In Sec.~III, using the same method, we 
derive jet broadening in DIS in terms of the same four-parton
correlation functions.  Finally, in Sec.~IV, we use the values of
$\lambda^2$ extracted from data on di-jet momentum imbalance and 
Drell-Yan $\Delta \langle q_T^2 \rangle$ to estimate  jet 
broadening in DIS.  We also present our discussions and 
conclusions in this section.

\section{nuclear enhancement for Drell-Yan 
$\langle q_T^2 \rangle$}

Consider the Drell-Yan process in hadron-nucleus collisions, 
$  h(p') + A(p) \rightarrow \ell^+\ell^-(q) + X $, where 
$q$ is the four-momentum for the virtual photon $\gamma^*$ 
which decays into the lepton pair.  $p'$ is the momentum for 
the incoming beam hadron and $p$ is the momentum per nucleon 
for the nucleus with the atomic number $A$. 

Let $q_T$ be the transverse momentum of the Drell-Yan pair, we
define the averaged transverse momentum square as
\begin{equation}
\langle q_T^2\rangle ^{hA}=
\left. \int dq_T^2 \cdot q_T^2 \cdot
\frac{d\sigma_{hA}}{dQ^2dq_T^2} \, \right/ 
\frac{d\sigma_{hA}}{dQ^2} \ .
\label{qt2}
\end{equation}
In Eq.~(\ref{qt2}), $Q$ is the total invariant mass of the lepton 
pair with $Q^2=q^2$. The transverse momentum spectrum, 
$d\sigma/dQ^2dq_T^2$, is sensitive to the multiple scattering,
and has the $A^{1/3}$-type nuclear size effect.  
If we write the cross section as 
\begin{equation}
\sigma_{hA} =  \sigma^S_{hA} + \sigma^D_{hA} + ... \ ,
\label{ea1}
\end{equation}
where $\sigma^S_{hA}$, $ \sigma^D_{hA}$, and ``...'' represent 
the single, double and higher multiple scattering, respectively. 
The single scattering is localized, hence, $\sigma_{hA}^S$ does not 
have a large dependence on the nuclear size.  Therefore, 
$\sigma_{hA}^S \approx A \sigma_{hN}$ with $N$ represents a nucleon. 
As we will demonstrate explicitly, the inclusive
cross section $d\sigma /dQ^2$ has much weaker nuclear 
dependence after integration over the whole momentum spectrum.
Hence,
\begin{equation}
\left. \int dq_T^2 \cdot q_T^2 \cdot 
\frac{d\sigma_{hA}^S}{dQ^2dq_T^2} \, \right/ 
\frac{d\sigma_{hA}}{dQ^2} \approx 
\langle q_T^2 \rangle ^{hN} \ .
\label{dy2}
\end{equation}
To extract effect due to multiple scattering we introduce
the nuclear enhancement of the Drell-Yan $\langle q_T^2 \rangle$ 
as 
\begin{equation}
\Delta \langle q_T^2\rangle 
\equiv \langle q_T^2 \rangle ^{hA}
      -\langle q_T^2 \rangle ^{hN} \ . 
\label{dydqt2}
\end{equation} 
If we keep only double scattering and neglect contribution from 
the higher multiple scattering, we have 
\begin{equation}
\Delta \langle q_T^2\rangle 
\approx 
\int d q_T^2 \cdot q_T^2\cdot 
\frac{d\sigma_{hA}^D}{dQ^2 dq_T^2}
\,    \left/ \frac{d\sigma_{hA}}{dQ^2} \right.  \ .
\label{dydqt2a}
\end{equation} 
From our definition, $\Delta \langle q_T^2\rangle$ represents a 
measurement of QCD dynamics beyond the traditional single-hard 
scattering picture.  The nuclear enhancement of Drell-Yan 
$\langle q_T^2 \rangle$, defined in Eq.~(\ref{dydqt2}), is a 
result of multiple scattering between the incoming beam parton 
and the nuclear matter before the Drell-Yan pair is produced. 

Since we are interested in the averaged transverse momentum square, 
we do not need to know the angular information of the Drell-Yan 
lepton pair.  After integration over the lepton's angular 
information, the Drell-Yan cross section can be written as 
\begin{equation}
d\sigma_{hA \rightarrow \ell^+ \ell^-}
= \left(\frac{2 \alpha_{em}}{3Q^2} \right) 
\, \left(-g_{\mu\nu}\, W^{\mu\nu}_{hA \rightarrow \gamma^*}(q) 
\right) \ ,
\label{dy3}
\end{equation}
where $W^{\mu\nu}$ is the hadronic tensor \cite{Fields}.
To simplify the notation of following derivation, we introduce
\begin{equation}
W_{hA\rightarrow\gamma^*}(q) \equiv
-g_{\mu\nu}\, W^{\mu\nu}_{hA\rightarrow \gamma^*}(q) \ .
\label{gW}
\end{equation}
At the leading order, the single scattering contribution to 
$W_{hA\rightarrow\gamma^*}(q)$ is given by
\begin{equation}
W^S_{hA\rightarrow\gamma^*}(q) 
= \sum_{q}\, 
\int\, dx'\, \phi_{\bar{q}/h}(x') \,
\int\, dx\, \phi_{q/A}(x) \, 
\left[\, e_q^2\, \frac{2\pi\alpha_{em}}{3}\, \right] \ ,
\label{bornW}
\end{equation}
where superscript ``S'' signals the single scattering;
and the corresponding inclusive single scattering cross section is
\cite{Fields}
\begin{equation}
\frac{d\sigma_{hA \rightarrow \ell^+ \ell^-}}{dQ^2} 
=\sigma_0 \, \sum_{q}\, e_q^2
\int\, dx'\, \phi_{\bar{q}/h}(x') \,
\int\, dx\, \phi_{q/A}(x) \, \delta (Q^2-xx's) \ , 
\label{dysingle}
\end{equation}
with $s=(p+p')^2$ and the Born cross section
\begin{equation}
\sigma_0=\frac{4\pi \alpha_{em}^2}{9Q^2} \ .
\label{sigma0}
\end{equation}

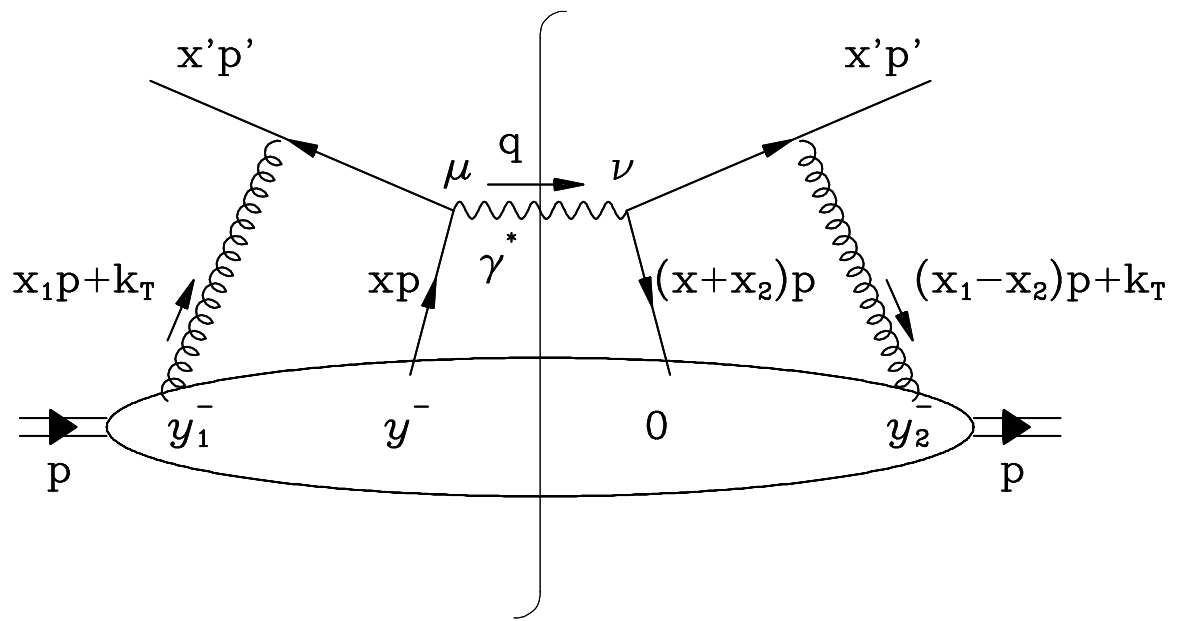
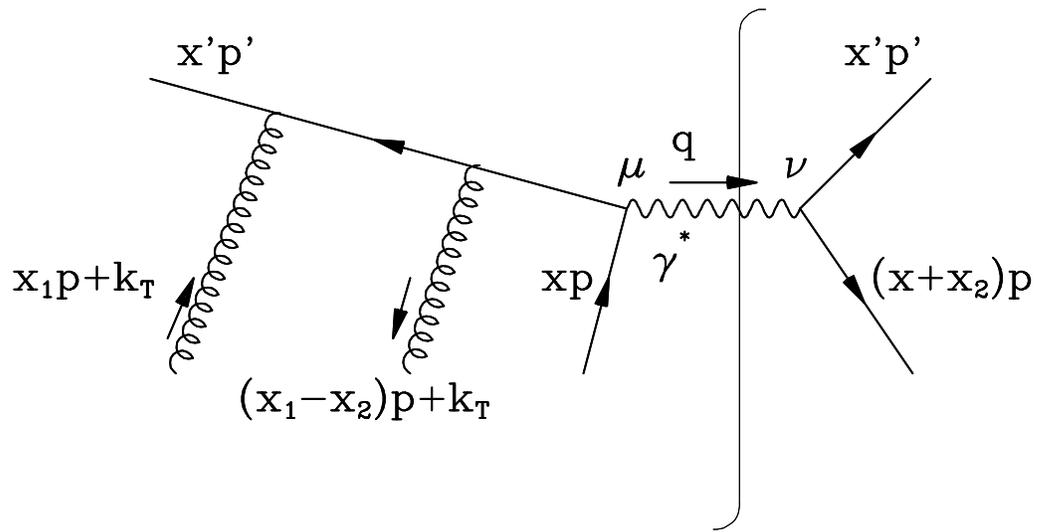
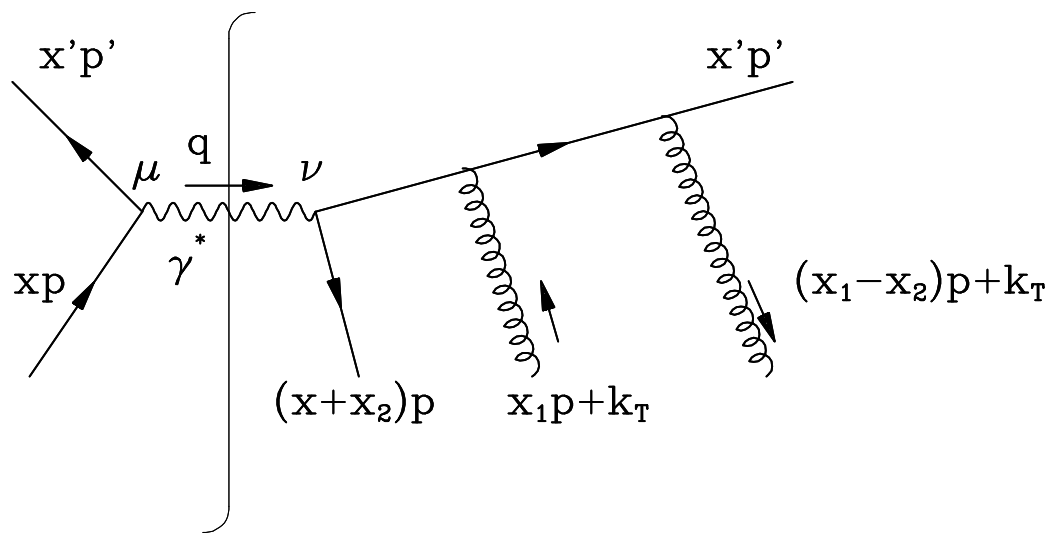
\begin{figure}
\caption{Lowest order double scattering contribution to 
the nuclear enhancement of Drell-Yan $\langle q_T^2 \rangle$; 
(a)symmetric diagram; (b) and (c): interference diagrams.} 
\label{fig1}
\end{figure}

Consider only double scattering inside the nuclear target,  
we can factorize the double scattering contribution to $W(q)$ as 
\begin{equation}
W^{D}_{hA\rightarrow\gamma^*}(q) 
= \sum_f\, \int dx'\, \phi_{f/h}(x') \cdot 
\hat{W}^{D}_{fA\rightarrow\gamma^*}(x',q)\ ,
\label{dydsigma}
\end{equation}
where superscript ``D'' indicates the double scattering 
contribution, and $f$ sums over all parton flavors.  In 
Eq.~(\ref{dydsigma}), $\hat{W}^{D}_{fA\rightarrow\gamma^*}(x',D)$ 
is the hadronic contribution from double scattering 
between a parton $f$ and the nucleus.  At the leading order in
$\alpha_s$, $\hat{W}^{D}_{fA\rightarrow\gamma^*}(x',D)$ is given by
the Feynman diagrams in Fig.~\ref{fig1}.  
According to the generalized factorization theorem \cite{QS},
$\hat{W}^{D}_{fA\rightarrow\gamma^*}(x',D)$ can be factorized as
\begin{equation}
\hat{W}^D_{bA \rightarrow \gamma^*}(x',D) = \frac{1}{2x's} \,
\int dx\, dx_1\, dx_2 \int d k_T^2 \, \bar{T}^{(I)}(x,x_1,x_2,k_T)\, 
\bar{H}(x,x_1,x_2,k_T,x'p',p,q) \, \Gamma^D \ ,
\label{dyD}
\end{equation}
where $\Gamma^D$ is the phase space factor for 
the double scattering. For diagram shown in Fig.~\ref{fig1}a,
for example, the phase space factor is given by 
\begin{eqnarray}
\Gamma^D_a &=& \frac{d^4q}{(2\pi)^4} \cdot (2\pi)^4
\delta^4 (xp+x_1p+x'p'+k_T-q)
\nonumber \\
&=&\frac{1}{x's}\, 
\delta (x+x_1-\frac{k_T^2}{x's}-\frac{Q^2}{x's}) \,
\delta (q_T^2-k_T^2) \,
dQ^2\, dq_T^2 \ .
\label{dyphase-a}
\end{eqnarray}
At this stage of the derivation, $\bar{T}$ and $\bar{H}$ in 
Eq.~(\ref{dyD}) both depend on gauge choice, even though 
$W(q)^D$ is gauge invariant.  In this paper, Feynman gauge 
is used in our calculation. 

In Eq.~(\ref{dyD}), the hadronic matrix element
\begin{eqnarray}
\bar{T}^{(I)}(x,x_1,x_2,k_T) &=&
\int \frac{dy^-}{2\pi} \,
\frac{dy_1^-}{2\pi}\, 
\frac{dy_2^-}{2\pi}\, 
\frac{d^2y_T}{(2\pi)^2} \,    \nonumber \\
&\ & \times \,
e^{ixp^+y^-}\, e^{ix_1p^+(y_1^--y_2^-)}\, 
e^{ix_2p^+y_2^-}\, e^{ik_T \cdot y_T}\,
\nonumber \\
&\  & \times \,
\frac{1}{2} \,\langle p_A |
A^+(y_{2}^{-},0_{T})\, \bar{\psi}_q(0)\, \gamma^+ \,
 \psi_q(y^{-}) \, A^+(y_{1}^{-},y_{1T}) | p_{A}\rangle \ .
\label{Tbari}
\end{eqnarray}
Because of the exponential factors in Eq.~(\ref{Tbari}), the position
space integration, $dy^-$'s cannot give large dependence on 
nuclear size unless the parton momentum fraction in one of 
the exponentials vanishes.  If the exponential vanishes, the corresponding
position space integration can be extended to the size of whole 
nucleus.  Therefore, in order to get large nuclear enhancement
or jet broadening, we need to consider only Feynman diagrams 
that can provide poles which set parton momentum fractions on
the exponentials to be zero \cite{LQS2}.  
At the leading order, only diagrams
shown in Fig.~\ref{fig1} have the necessary poles.
These diagrams contribute to the double scattering partonic
part $\bar{H}(x,x_1,x_2,k_T,x'p',p,q)$ in Eq.~(\ref{dyD}).

For the leading order diagrams shown in Fig.~\ref{fig1}, the 
corresponding partonic parts have two possible poles.  For example,
for diagram shown in Fig.~\ref{fig1}, the partonic part has following
general structure
\begin{eqnarray}
\bar{H}^a
& \sim & 
\frac{1}{(x'p'+x_1p+k_T)^2 +i\epsilon} \cdot 
\frac{1}{(x'p'+(x_1-x_2)p+k_T)^2-i\epsilon} \cdot 
(-g_ {\mu \nu} ) \, \hat{H}^{\mu\nu}
\nonumber \\
&\propto &  \frac{1}{x_1-\frac{k_T^2}{x's}+i\epsilon}
\cdot \frac{1}{x_1-x_2-\frac{k_T^2}{x's}-i\epsilon} \ ,
\label{dypole-a}
\end{eqnarray}
where $\hat{H}^{\mu\nu}$ is the numerator which is proportional to
a quark spinor trace.

The first step to evaluate the Eq.~(\ref{dyD}) is to carry out the
integration over parton momentum fractions $dx\, dx_1\, dx_2$.
Using one of the $\delta$-functions in phase space factor
in Eq.~(\ref{dyphase-a}) to fix $dx$ integration, and the two poles 
in the partonic part in Eq.~(\ref{dypole-a}) to perform the contour 
integration for $dx_1 \, dx_2$, we obtain
\begin{eqnarray}
I_a &\equiv & \int dx\, dx_1\, dx_2\, \
e^{ixp^+y^-} \, e^{ix_1p^+(y_1^--y_2^-)}\, 
e^{ix_2p^+y_2^-}\,
\nonumber \\
&\ & \times\, 
\frac{1}{x_1-\frac{k_T^2}{x's}+i\epsilon}
\cdot \frac{1}{x_1-x_2-\frac{k_T^2}{x's}-i\epsilon}
\cdot \delta (x+x_1-\frac{k_T^2}{x's}-\frac{Q^2}{x's})
\nonumber \\
&=& (4\pi^2)\,\theta (y^--y_1^-) \,\theta (-y_2^-)\, 
e^{i \, (\tau/x')\, p^+ y^-} \, 
e^{i \, (k_T^2/(x's))\, p^+ (y_1^--y_2^-)} \ , 
\label{dyIa}
\end{eqnarray}
with $\tau=Q^2/s$.  For the diagram shown in 
Fig.~\ref{fig1}b, we have a slightly different phase space factor
\begin{equation}
\Gamma^D_b = \frac{1}{x's}\,
\delta (x+x_2-\frac{Q^2}{x's}) \, 
\delta (q_T^2) \, 
dQ^2\, dq_T^2
\ .
\label{dyphase-b}
\end{equation}
The corresponding partonic part $\bar{H}^b$ has slightly different 
poles 
\begin{equation}
\bar{H}^b
\propto   \frac{1}{x_1-\frac{k_T^2}{x's}+i\epsilon}
\cdot \frac{1}{x_2+i\epsilon} \ .
\label{dypole-b}
\end{equation}
Integrating over parton momentum fractions, $dx\, dx_1\, dx_2$, 
we have 
\begin{eqnarray} 
I_b &\equiv & \int dx\, dx_1\, dx_2\, 
e^{ixp^+y^-}\, e^{ix_1p^+(y_1^--y_2^-)}\, 
e^{ix_2p^+y_2^-}\,
\nonumber \\
&\ & \times \,
\frac{1}{x_1-\frac{k_T^2}{x's}+i\epsilon}
\cdot \frac{1}{x_2+i\epsilon} \cdot \delta (x+x_2-\frac{Q^2}{x's})
\nonumber \\
&=& (-4\pi^2)\, \theta (y_2^--y_1^-) \,
\theta (y^--y_2^-) \, e^{i\, (\tau/x') p^+ y^- } \, 
e^{i\, (k_T^2/(x's)) p^+ (y_1^--y_2^-)} \ .
\label{dyIb}
\end{eqnarray}
Similarly, for diagram shown in Fig~\ref{fig1}c, we have  
\begin{eqnarray}
I_c =  (-4\pi^2)\, \theta (y_1^--y_2^-) \,
\theta  (-y_1^-) \,  e^{i\, (\tau/x') p^+ y^-} \, 
e^{i (k_T^2/(x's)) p^+ (y_1^--y_2^-)} \ .
\label{dyc}
\end{eqnarray}

It is easy to show that the numerator spinor trace gives the same 
factor for all three diagrams. Therefore, total contribution to
jet broadening is proportional to $I_a+I_b+I_c$, and this sum
has the following feature
\begin{eqnarray}
\frac{d\sigma^D_{hA}}{dQ^2\,dq_T^2}
& \propto &
I_a+I_b+I_c  \\
&\propto  &  \theta (y^--y_1^-) \, 
\theta (-y_2^-) \, \left[ \delta (q_T^2-k_T^2) - \delta (q_T^2) \right] 
\nonumber \\
&+& \left[
\theta (y^--y_1^-) \,\theta (-y_2^-) -
\theta (y_2^--y_1^-)\, \theta (y^--y_2^-) -
\theta (y_1^--y_2^-)\, \theta (-y_1^-) \right]\,
\delta (q_T^2) \ .
\label{Isum}
\end{eqnarray} 
It is clear from Eq.~(\ref{Isum}) that for the inclusive Drell-Yan cross 
section, $d\sigma/dQ^2$, the double scattering contribution, 
$d\sigma^D_{hA}/dQ^2$, does not have large dependence on the nuclear size.
The integration over $dq_T^2$ eliminates the first term in Eq.~(\ref{Isum}),
while the second term is localized in space if $\tau/x'$ is not too 
small.  When the $\tau/x'$ is finite, and $p^+$ is large, 
the exp$[i\,(\tau/x') p^+y^-]$  effectively
restricts $y^{-}\sim 1/((\tau/x') p^+) \rightarrow 0$.  
When $y^-\rightarrow 0$, the combination of 
three pairs of $\theta$-functions in Eq.~(\ref{Isum}) vanishes,
\begin{eqnarray}
\frac{d\sigma^D_{hA}}{dQ^2} &\equiv &
\int dq_T^2 \left( \frac{d\sigma^D_{hA}}{dQ^2dq_T^2} \right)
\nonumber \\
&\propto &
\left[
\theta (y^--y_1^-) \,\theta (-y_2^-) -
\theta (y_2^--y_1^-)\, \theta (y^--y_2^-) -
\theta (y_1^--y_2^-)\, \theta (-y_1^-) \right]
\nonumber \\
& \rightarrow & 0 \quad\quad \mbox{as $y\rightarrow 0$} \ .
\label{totalD}
\end{eqnarray}
Physically, Eq.~(\ref{totalD}) says that 
all integrations of $y^-$'s are localized.
Actually, at the leading order, the term proportional to 
$\theta (y^--y_1^-) \,\theta (-y_2^-) -
\theta (y_2^--y_1^-)\, \theta (y^--y_2^-) -
\theta (y_1^--y_2^-)\, \theta (-y_1^-)$ is the eikonal contribution 
in Feynman gauge to make the normal twist-2 quark distribution
$\phi_{q/A}(x)$ in Eq.~(\ref{dysingle}) gauge invariant.
Eq.~(\ref{totalD}) is a good example to demonstrate that 
the double scattering does not give large nuclear size effect 
to the total inclusive cross section.

On the other hand, from Eq.~(\ref{Isum}),  
the double scattering contribution to the 
averaged transverse momentum square can gain large size effect
\begin{eqnarray}
\Delta \langle q_T^2 \rangle 
& \sim &
\int dq_T^2 \, q_T^2\, 
\left( \frac{d\sigma^D_{hA}}{dQ^2dq_T^2} \right)
\nonumber \\
&\propto &
\int dq_T^2 \, q_T^2\,
\left[ \delta (q_T^2-k_T^2) - \delta (q_T^2) \right] 
\nonumber \\
&\sim & k_T^2 \ .
\label{qt2Da}
\end{eqnarray}
Actually, $k_T^2$ in Eq.~(\ref{qt2Da}) needs to be integrated first. But 
Eq.~(\ref{qt2Da}) already demonstrates that 
$\Delta \langle q_T^2 \rangle$ is proportional 
$k_T^2$, which is the kick of the transverse momentum the parton received 
from the additional scattering. The bigger the nuclear size, the bigger 
effective $k_T^2$. As shown below,  $\Delta \langle q_T^2 \rangle$ is 
proportional to the nuclear size.

After demonstrating the physical picture of transverse momentum 
broadening in the Drell-Yan process, 
we are now ready to carry out the algebra.  
Working out the spinor trace for all three diagrams in Fig.~\ref{fig1}, 
and combining three diagrams and drop the term localized in space, 
we obtain
\begin{eqnarray}
\Delta \langle q_T^2 \rangle\, \frac{d\sigma_{hA}}{dQ^2}
&=&
\int dq_T^2 \, q_T^2\ \sigma_0 \,
\sum_{q} \int\, dx'\, \phi_{\bar{q}/h}(x') 
\int\, dx\, \delta (Q^2-xx's) 
\nonumber \\
&\ & \times 
\int \frac{dy}{2\pi}\, \frac{dy_1dy_2}{2\pi}\,
\theta (y^--y_1^-) \, 
\theta (-y_2^-) \, e^{i x p^+ y^- }
\nonumber \\
&\ & \times 
\frac{d^2y_T}{(2\pi)^2} \,
\int d^2 k_T\,  e^{ik_T \cdot y_T} \, 
e^{i (k_T^2/(x's)) p^+ (y_1^--y_2^-)} \,
\nonumber \\ 
&\ & \times\,
\frac{1}{2}\,
\langle p_A |
A^+(y_{2}^{-},0_{T})\, \bar{\psi}_q(0)\, \gamma^+ \,
            \psi_q(y^{-})\, A^+(y_{1}^{-},y_{1T}) | p_{A}\rangle 
\nonumber \\
&\ & \times  
\left( \frac{4\pi^2 \alpha_s}{3} \right) 
\cdot  \left[ \delta (q_T^2-k_T^2) - \delta (q_T^2) \right] 
\label{qt2Dbb} \\
& \approx &
\sigma_0  \left(\frac{4\pi^2 \alpha_s}{3} \right) 
\sum_{q} \int\, dx'\, \phi_{\bar{q}/h}(x') 
\int\, dx\, T_{q/A}^{(I)}(x)\, \delta (Q^2-xx's) \ ,
\label{qt2Db}
\end{eqnarray}
where $\sigma_0$ is defined in Eq.~(\ref{sigma0}) and 
the four-parton correlation function is given by
\begin{eqnarray}
T_{q/A}^{(I)}(x) &=&
 \int \frac{dy^{-}}{2\pi}\, e^{ixp^{+}y^{-}}
 \int \frac{dy_1^{-}dy_{2}^{-}}{2\pi} \,
      \theta(y^{-}-y_{1}^{-})\,\theta(-y_{2}^{-}) \nonumber \\
&\ & \times \,
     \frac{1}{2}\,
     \langle p_{A}|F_{\alpha}^{\ +}(y_{2}^{-})\bar{\psi}_{q}(0)
                  \gamma^{+}\psi_{q}(y^{-})F^{+\alpha}(y_{1}^{-})
     |p_{A} \rangle \ .
\label{dyTq}
\end{eqnarray}
In deriving Eq.~(\ref{qt2Db}), we expend $\delta (q_T^2-k_T^2)$ 
at $k_T=0$, known as the collinear expansion, and keep only the first 
non-vanishing term which corresponds to 
the second order derivative term,
\begin{equation}
\delta (q_T^2-k_T^2) - \delta (q_T^2) 
\approx 
-\delta' (q_T^2) (-g^{\perp}_{\alpha \beta})\, 
k_T^{\alpha}\, k_T^{\beta} \ .
\label{dy7}
\end{equation}
We use the factor $k_T^{\alpha} k_T^{\beta}$ in Eq.~(\ref{dy7}) 
to convert the $k_T^{\alpha}A^+k_T^{\beta}A^+$ into field strength 
$F^{\alpha+}F^{\beta+}$ by partial integration. 
Here, we work in Feynman gauge. The terms associated with 
other components of $A^\rho$ 
are suppressed by $1/p^+$ compared to those with 
$A^+$, because of the requirement 
of the Lorentz boost invariance for the matrix elements \cite{LQS1}.
$T_{q/A}^{I}(x)$ defined in Eq.~(\ref{dyTq}) is the four-parton 
correlation function in a nucleus. The superscript (I) represents the
initial state interaction, in order to distinguish from the 
similar four-parton correlation function defined in Eq.~(\ref{Ta}). 
More discussion on the relation between $T_{q/A}(x)$ and 
$T_{q/A}^{(I)}(x)$ will be given in Sec.~IV.

Combining Eqs.~(\ref{dydqt2}), (\ref{dysingle}), and (\ref{qt2Db}), we 
obtain the nuclear enhancement of Drell-Yan $\langle q_T^2 \rangle$ 
at the leading order in $\alpha_s$
\begin{equation}
\Delta \langle q_T^2 \rangle  
=\left(\frac{4\pi^2 \alpha_s}{3} \right)\cdot
\frac{\sum_{q} \, e_q^2\int dx' \, \phi_{\bar{q}/h}(x')\, 
T^{(I)}_{q/A}(\tau /x') /x'}
{\sum_{q}\, e_q^2 \int dx' \, \phi_{\bar{q}/h}(x')\, 
\phi_{q/A}(\tau /x) /x'} \ .
\label{dyqt2b}
\end{equation}

\section{Jet broadening in Deeply Inelastic Scattering}

Consider the jet production in  the deeply inelastic lepton-nucleus 
scattering, $e(k_1) + A(p) \rightarrow e(k_2) +jet(l) +X$. 
$k_1$ and $k_2$ are the four
momentum of the incoming and outgoing lepton respectively,  
and $p$ is the momentum per nucleon for the nucleus with the atomic 
number $A$. 
With $l$ being the observed jet momentum, we define the averaged 
jet transverse momentum square as
\begin{equation}
\langle l_T^2\rangle^{eA} =\left. \int dl_T^2 \cdot l_T^2 \cdot  
\frac{d\sigma_{eA}}{dx_BdQ^2dl_T^2} \, 
\right/ \frac{d\sigma_{eA}}{dx_BdQ^2}
\ ,
\label{lt2}
\end{equation}
where $x_B=Q^2/(2p\cdot q)$, 
$q=k_1-k_2$ is the momentum of the virtual photon, and $Q^2=-q^2$. 
The jet transverse momentum $l_T$ depends on our choice of the frame. 
We choose the Breit frame in the following calculation. 
Similar to Drell-Yan $d\sigma/dQ^2 dq_T^2$, the jet transverse 
momentum spectrum, $d\sigma/dx_BdQ^2dl_T^2$, is sensitive to 
the $A^{1/3}$ type nuclear size effect due to multiple scattering. 
On the other hand, the inclusive DIS cross section 
$d\sigma/dx_BdQ^2 = \int d l_T^2\, d\sigma/dx_BdQ^2dl_T^2$ 
does not have the $A^{1/3}$ power enhancement.  Instead, it has 
the much weaker A-dependence, such as the EMC effect and nuclear 
shadowing.  To separate the multiple scattering contribution from 
the single scattering, we define the jet broadening as 
\begin{equation}
\Delta \langle l_T^2\rangle  \equiv \langle l_T^2\rangle ^{eA} -
\langle l_T^2 \rangle ^{eN} \ .
\label{broaden} 
\end{equation}
Keeping only contribution from the double scattering, 
similar to Eq.~(\ref{dydqt2}), we have 
\begin{equation}
\Delta \langle l_T^2\rangle
\approx
\left. \int dl_T^2 \cdot l_T^2 \cdot 
\frac{d\sigma^{D}_{eA}}{dx_BdQ^2dl_T^2} \, \right/ 
\frac{d\sigma_{eA}}{dx_BdQ^2} \ .
\label{dlt2d}
\end{equation} 
In the rest of this section, we derive the leading contribution to
$\Delta \langle l_T^2 \rangle$ by using the same technique 
used to derive the Drell-Yan $\Delta \langle q_T^2 \rangle$ 
in last section.

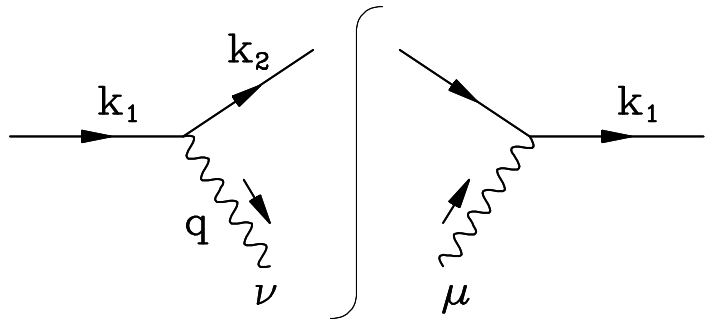
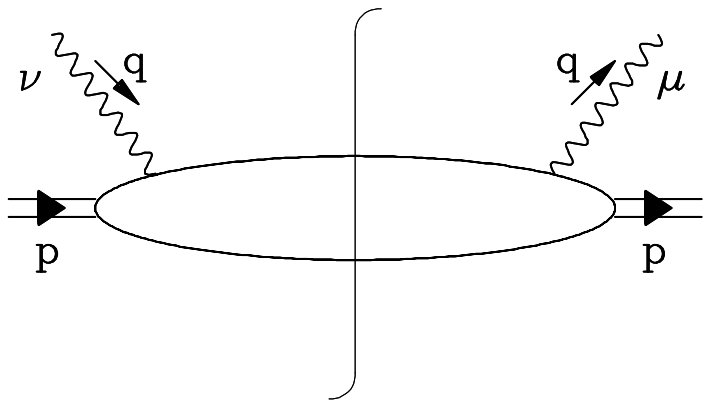
\begin{figure}
\caption{ Diagrams for DIS: (a) Diagram representing $L_{\mu\nu}$; 
(b) Diagram representing $W^{\mu\nu}$.}
\label{fig2}
\end{figure}

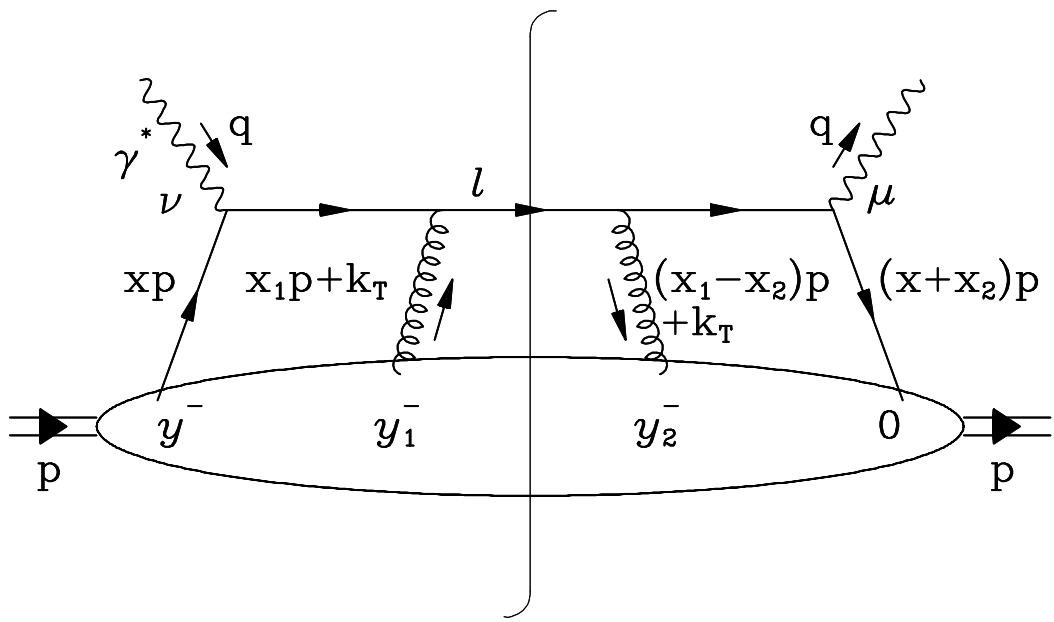
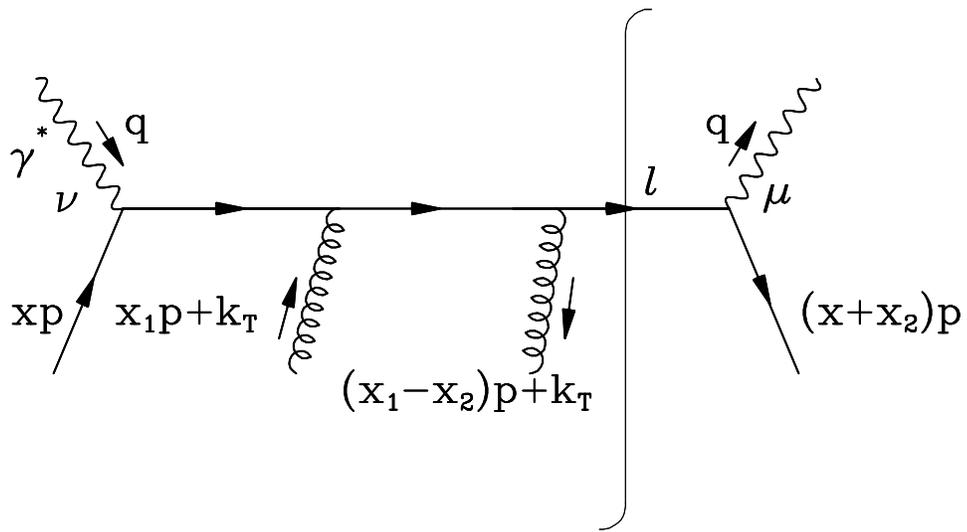
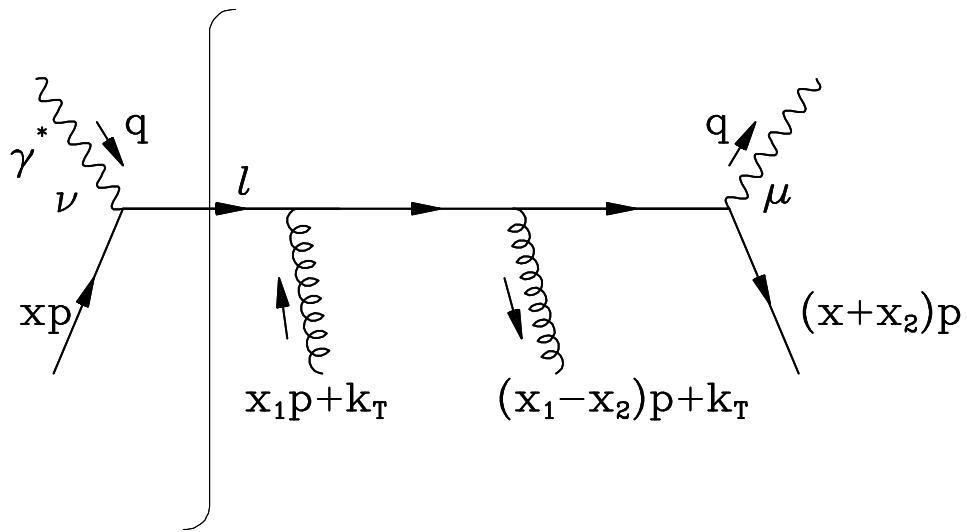
\begin{figure}
\caption{Lowest order double scattering contribution to jet broadening: 
(a) symmetric diagram; (b) and (c): interference diagrams.}
\label{fig3}
\end{figure}

The general expression for the cross section in DIS is 
\begin{equation}
d\sigma=\frac{1}{2s}\,\frac{e^2}{Q^4}\,
L_{\mu\nu}\,W^{\mu\nu}\, 
\frac{d^3k_2}{(2\pi)^3 \, 2E_2} \ ,
\label{sigma}
\end{equation}
with $s=(p+k_1)^2$.
In Eq.~(\ref{sigma}), the leptonic tensor $L_{\mu\nu}$ is 
given by the diagram in Fig.~\ref{fig2}a,  
\begin{equation}
L_{\mu\nu}=\frac{1}{2} {\rm Tr}(\gamma \cdot k_1 \gamma_{\mu}
\gamma \cdot k_2 \gamma_{\nu}) \ ,
\label{L}
\end{equation}
and $W^{\mu\nu}$ is the hadronic tensor given by the diagram 
shown in Fig.~\ref{fig2}b.  The leading order double scattering
diagrams contributing to jet broadening are given in
Fig.~\ref{fig3}.  

Feynman diagrams in Fig.~\ref{fig3} give the double scattering 
contribution to the hadronic tensor $W_{\mu\nu}$ as, 
\begin{equation}
W_{\mu\nu}^D = \sum_a \,
\int dx\, dx_1\, dx_2 \int d k_T^2 \, 
\bar{T}^{(F)}(x,x_1,x_2,k_T) 
\bar{H}_{\mu\nu}(x,x_1,x_2,k_T,p,q,l) \ ,
\label{W1}
\end{equation}
with the matrix element 
\begin{eqnarray}
\bar{T}^{(F)}(x,x_1,x_2,k_T)  &=&
\int \frac{dy^-}{2\pi}\, \frac{dy_1^-}{2\pi}\, 
\frac{dy_2^-}{2\pi} \,
\frac{d^2y_T}{(2\pi)^2} \,    \nonumber \\
&\ &\, \times e^{ixp^+y^-}\, e^{ix_1p^+(y_1^--y_2^-)}\, 
e^{ix_2p^+y_2^-}\, e^{ik_T \cdot y_T}
\nonumber \\
&\  &\, \times \frac{1}{2} \,\langle p_A |
 \bar{\psi}_q(0)\, \gamma^+ \,A^+(y_{2}^{-},0_{T})\,
          A^+(y_{1}^{-},y_{1T})\, \psi_q(y^{-}) | p_{A}\rangle \ ,
\label{TbarF}
\end{eqnarray}
where superscript ($F$) indicates the final-state double scattering.
The matrix element $\bar{T}^{(F)}$ is equal to the  
$\bar{T}^{(I)}$ in Eq.~(\ref{Tbari}) if we commute the gluon fields 
with the quark fields \cite{QS,jaffe}.
In Eq.~(\ref{W1}), $\bar{H}_{\mu\nu}(x,x_1,x_2,k_T,p,q,l)$ is 
the corresponding partonic part. 

For the diagram shown in Fig.\ref{fig3}a, the partonic part is given
by
\begin{eqnarray}
\bar{H}^a_{\mu\nu} &=& C\cdot \hat{H}_{\mu\nu}\cdot 
\frac{1}{(xp+q)^2+i\epsilon} \cdot 
\frac{1}{(xp+x_2p+q)^2-i\epsilon} 
\nonumber \\
&\ & \times (2\pi)\, \delta ((xp+x_1p+k_T+q)^2)\,
\delta (l_T^2 -k_T^2) \, dl_T^2  
\nonumber \\ 
&=& C \cdot \hat{H}_{\mu\nu} \cdot \frac{2\pi}{(2p\cdot q)^3} \cdot
\delta (l_T^2 -k_T^2) \, dl_T^2  
\nonumber \\
&\ & \times \delta (x+x_1-x_B -\frac{k_T^2-2q\cdot k_T}{2p\cdot q}) \cdot
\frac{1}{x-x_B+i\epsilon} \cdot 
\frac{1}{x+x_2-x_B-i\epsilon} \ ,
\label{Hbar}
\end{eqnarray}
where $C$ is the color factor for the diagrams in Fig.~\ref{fig3}, and 
given by $C=1/6$.
In Eq.~(\ref{Hbar}), the $\delta$-function is from the phase space, and 
$\hat{H}_{\mu\nu}$ represents the spinor trace and is given by 
\begin{equation}
\hat{H}_{\mu\nu} =(4\pi)^2 \alpha_s \alpha_{em} e_q^2 \,
{\rm Tr}\left[\gamma \cdot p \gamma_{\mu} \gamma \cdot (xp+x_2p+q) 
\gamma^{\sigma} \gamma \cdot l \gamma^{\rho} \gamma \cdot (xp+q)
\gamma_{\nu}\right] \, p_{\rho}\,p_{\sigma}\ ,
\label{hatH}
\end{equation}
where $p_{\rho}\,p_{\sigma}$ is a result of our definition for 
the hadronic matrix element in Eq.~(\ref{TbarF}).

Following the derivation of the Drell-Yan $\langle q_T^2 \rangle$, first, 
we carry out the integrations of the parton momentum fractions by
using the $\delta$-function and two poles in Eq.~(\ref{Hbar}),
\begin{eqnarray}
H^a &\equiv & 
\int dx\, dx_1\, dx_2 \, e^{ixp^+y^-} \, 
e^{ix_1p^+(y_1^--y_2^-)}\, e^{ix_2p^+y_2^-}\, 
\delta(l_T^2-k_T^2) 
\nonumber \\
&& \times \delta (x+x_1-x_B -\frac{k_T^2-2q\cdot k_T}{2p\cdot q}) 
\cdot \frac{1}{x-x_B+i\epsilon} \cdot 
\frac{1}{x+x_2-x_B-i\epsilon} \nonumber \\
&=& (2\pi)^2\, \theta (y_1^--y^-) \,\theta (y_2^-)\, e^{ix_Bp^+y^-} \, 
\delta(l_T^2-k_T^2)\, 
e^{i\left(\frac{k_T^2-2q\cdot k_T}{2p\cdot q}\right) 
    \, p^+ (y_1^--y_2^-)} \ .
\label{intHa}
\end{eqnarray}
Similarly, we have the corresponding integrations for the interference 
diagram shown in Fig.~\ref{fig3}b
\begin{eqnarray}
H^b &\equiv & 
\int dx\, dx_1\, dx_2 \, e^{ixp^+y^-} \, 
e^{ix_1p^+(y_1^--y_2^-)}\, e^{ix_2p^+y_2^-}\, 
\delta(l_T^2)  \nonumber \\
&& \times \delta (x+x_1-x_B) \cdot 
\frac{1}{x-x_B+i\epsilon} \cdot 
\frac{1}{x+x_1-x_B-\frac{k_T^2-2q\cdot k_T}{2p\cdot q}+i\epsilon} 
\nonumber \\
&=& - (2\pi)^2\, \theta (y_2^--y_1^-) \,\theta (y_1^--y^-)\, 
e^{ix_Bp^+y^-} \, \delta(l_T^2) \,
e^{i\left(\frac{k_T^2-2q\cdot k_T}{2p\cdot q}\right) 
    \, p^+ (y_1^--y_2^-)} \ ,
\label{intHb}
\end{eqnarray}
and for the diagram in Fig.~\ref{fig3}c
\begin{eqnarray}
H^c &\equiv & 
\int dx\, dx_1\, dx_2 \, e^{ixp^+y^-} \, 
e^{ix_1p^+(y_1^--y_2^-)}\, e^{ix_2p^+y_2^-}\, 
\delta(l_T^2)   \nonumber \\
&& \times \delta (x-x_B) \cdot 
\frac{1}{x+x_1-x_B-\frac{k_T^2-2q\cdot k_T}{2p\cdot q}-i\epsilon} \cdot 
\frac{1}{x+x_2-x_B-i\epsilon} \nonumber \\
&=& -(2\pi)^2\, \theta (y_2^-) \,\theta (y_1^--y_2^-)\, 
e^{ix_Bp^+y^-} \, \delta(l_T^2) \,
e^{i\left(\frac{k_T^2-2q\cdot k_T}{2p\cdot q}\right) 
    \, p^+ (y_1^--y_2^-)} \ .
\label{intHc}
\end{eqnarray}
Combining Eqs.~(\ref{intHa}), (\ref{intHb}) and (\ref{intHc}), we 
again have the same structure as that in Eq.~(\ref{Isum}).  Therefore, 
like the Drell-Yan case, we conclude that the 
double scattering does not result 
into any large nuclear size dependence to the inclusive DIS cross section. 
For the jet broadening defined in Eq.~(\ref{dlt2d}), we drop 
the term proportional to $\theta (y_1^--y^-) \,\theta (y_2^-) -
\theta (y_2^--y_1^-)\, \theta (y_1^--y^-) -
\theta (y_1^--y_2^-)\, \theta (y_2^-)$, which is 
localized as the single scattering.  After carrying out
the algebra similar to those following Eq.~(\ref{qt2Da}), we derive 
the jet broadening in DIS as
\begin{equation}
\Delta \langle l_T^2 \rangle 
= \left(\frac{4\pi^2 \alpha_s}{3} \right)
\cdot \frac{\sum_{q}\, e_q^2\, T_{q/A}(x_B)}
           {\sum_{q}\, e_q^2\, \phi_{q/A}(x_B)} \ ,
\label{lt2T}
\end{equation}
where $\sum_q$ sums over all quark and antiquark flavors.
In Eq.~(\ref{lt2T}), $\phi_{q/A}(x)$ is the normal twist-2 quark 
distribution inside a nucleus, and the four-parton correlation
function, $T_{q/A}(x_B)$ is defined in Eq.~(\ref{Ta}).  

From Eq.~(\ref{lt2T}), we conclude that based on the leading order
calculation, we can uniquely predict the jet broadening in DIS,
without any free parameter, if the four-parton correlation function,
$T_{q/A}(x_B)$, is measured in another experiment.  Therefore,
jet broadening in DIS can be used to test the QCD treatment of
multiple scattering in nuclear environment.  In addition, the jet 
broadening given in Eq.~(\ref{lt2T}) is directly proportional to
$T_{q/A}(x_B)$, $x_B$-dependence of $\Delta \langle l_T^2 \rangle$ 
can provide immediate information on the functional form of the 
four-parton correlation function.


\section{Discussion and conclusions}

From Eqs.~(\ref{dyqt2b}) and (\ref{lt2T}), both jet broadening in DIS 
and the nuclear enhancement of the Drell-Yan $\langle q_T^2\rangle$ are 
directly proportional to the four-parton (twist-4) correlation 
functions, which are defined in Eqs.~(\ref{Ta}) and (\ref{dyTq}), 
respectively.  Since field operators are commuted on the light-cone 
\cite{QS,jaffe}, the four-parton correlation functions,
$T_{q/A}(x)$ and $T^I_{q/A}(x)$, are the same if the phase space 
integral are symmetric.  With the model given in Eq.~(\ref{TiM}),
we predict the jet broadening in DIS as
\begin{equation}
\Delta \langle l_T^2 \rangle = 
\left(\frac{4\pi^2\alpha_s}{3} \right) \
\lambda^2\ A^{1/3} \ ;
\label{dlt2}
\end{equation}
and predict the nuclear enhancement of Drell-Yan transverse momentum
square as 
\begin{equation}
\Delta \langle q_T^2 \rangle = 
\left(\frac{4\pi^2\alpha_s}{3} \right) \
\lambda^2\ A^{1/3} \ .
\label{dyqt2c}
\end{equation} 
which is the same as that obtained in Ref.~\cite{YuriD}.
It is a direct result of the model proposed by LQS \cite{LQS2}
that both jet broadening in DIS and nuclear enhancement in 
Drell-Yan $\Delta \langle q_T^2 \rangle$ have the very simple
expressions, shown in Eqs.~(\ref{dlt2}) and (\ref{dyqt2c}).
In addition, Eqs.~(\ref{dlt2}) and (\ref{dyqt2c})
tell us that at the leading 
order, jet broadening in DIS and nuclear enhancement in Drell-Yan 
$\Delta \langle q_T^2 \rangle$ 
are the same, if the averaged 
initial-state gluon interactions is equal to the corresponding 
final-state gluon interactions, i.e., 
$T_{q/A}(x)=T_{q/A}^I(x)$.  

From the simple expression in Eq.~(\ref{dlt2}), we conclude
that at the leading order, the jet broadening in DIS has a 
strong scaling property, and it does not depend on beam energy, 
$Q^2$ and $x_B$.  However, 
the $x_B$-dependence needs to be modified if
$x_B$ is smaller than 0.1.  When $x_B \leq 0.1$, the 
localized term, $\theta (y_1^--y^-) \,\theta (y_2^-) -
\theta (y_2^--y_1^-)\, \theta (y_1^--y^-) -
\theta (y_1^--y_2^-)\, \theta (y_2^-)$, is no longer localized 
\cite{strikman,MQ}.  Therefore, we have to keep this term 
for jet broadening calculation if $x_B$ is small.  
In addition, $Q^2$-dependence may be modified because 
the four-parton correlation function
$T_{q/A}(x)$ and the normal quark distribution $\phi_{q/A}(x)$
can have different scaling violation.
Of course, all dependence or whole conclusion could be modified due
to possible different high order corrections.  Nevertheless, we
believe that experimental measurements of the jet broadening in DIS 
can provide valuable information on strength of multi-parton 
correlations and dynamics of multiple scattering.
 
Similarly, from Eq.~(\ref{dyqt2c}), we can also conclude that 
nuclear enhancement in the Drell-Yan $\langle q_T^2 \rangle$ has 
small dependence on beam energy and $Q^2$ of the lepton pair.
However, data from Fermilab E772 and CERN NA10 \cite{E772,NA10}
demonstrate some energy dependence.  It signals that high order
corrections for Drell-Yan $\langle q_T^2 \rangle$ is important
\cite{Guo,GQS}, or the simple model of LQS for four-parton 
correlation functions is too simple.  On the other hand, the 
weak energy dependence from the data shows that the leading 
order calculation given here can be useful, and QCD treatment
of multiple scattering can be eventually tested and understood.

Use Eq.~(\ref{dyqt2c}) and data from E772 and NA10 on nuclear 
enhancement of Drell-Yan $\langle q_T^2 \rangle$, we estimate 
the value of $\lambda^2$ as
\begin{equation}
\lambda^2_{\mbox{\tiny DY}} \approx 0.01 \mbox{GeV}^2 \ ,
\label{dylambda}
\end{equation}
which is at least a factor of five smaller than 
$\lambda^2_{\mbox{\tiny di-jet}} \approx 0.05\sim 0.1$ GeV$^2$,
estimated by LQS from momentum imbalance of di-jet data 
\cite{LQS2}.

Since jet broadening in DIS and the momentum imbalance in di-jet 
are both due to the final-state multiple scattering, we will use
the value of $\lambda^2_{\mbox{\tiny ji-jet}}$ to predict
the jet broadening as
\begin{equation}
\Delta \langle l_T^2 \rangle \approx (0.66 \sim 1.31) \,
\alpha_s \, A^{1/3} \ .
\label{dijet}
\end{equation}
On the other hand, with a smaller value of 
$\lambda^2_{\mbox{\tiny DY}}$ in Eq.~(\ref{dylambda}), we will predict
a much smaller jet broadening $\sim 0.13\, \alpha_s \, A^{1/3}$.
Direct experimental measurement on jet broadening can certainly
provide an independent test of our QCD treatment of multiple scattering.
Early data on jet production from Fermilab E665 can be used to
calculate the jet broadening.  Future experiments at HERA with a 
heavy ion beam \cite{strikman} should be able to provide much more 
information on dynamics of parton correlations.

In summary, the jet broadening in DIS provides an independent 
measurement of four-parton correlation functions, and tests of 
QCD dynamics beyond simple parton model.  At the leading order, 
jet broadening is  directly proportional to the four-parton 
correlation functions. 
The size of jet broadening provides a direct measurement of the 
size of the four-parton correlation functions. 
In addition, by varying the $x_B$, we can gain valuable information 
on $x$-dependence of the four-parton correlation functions.
Information on $x_B$-dependence of the jet broadening, $\Delta
\langle l_T^2 \rangle$ in Eq.~(\ref{lt2T}), 
can provide a direct measurement of
the functional form of $T_{q/A}(x_B,A)$ defined in Eq.~(\ref{Ta}).


\end{document}